\pdfoutput=1
\documentclass[12pt]{article}

\usepackage[margin=1in]{geometry}

\usepackage{graphicx}
\usepackage{url}
\usepackage{soul}
\usepackage{wrapfig}
\usepackage{lscape}
\usepackage{rotating}
\usepackage{epstopdf}
\usepackage{authblk}
\usepackage{setspace}
\usepackage{hyperref}
\usepackage{hyperref}

\usepackage[authoryear]{natbib}

\usepackage{authblk}

\begin{document}

\begin{center}

{\bfseries\fontsize{14}{16}\selectfont
Storm-Time Cusp Precipitation: Insights from TRACERS Multi-Crossing Observations
\par}

\vspace{0.4cm}

{\fontsize{12}{14}\selectfont
Shirsh Lata Soni$^{1}$, David M. Miles$^{1}$, Jasper Halekas$^{1}$,
Stephen A. Fuselier$^{2,3}$, George Hospodarsky$^{1}$,
Marit Øieroset$^{4}$, K. J. Trattner$^{2}$, S. M. Petrinec$^{2}$,
Jeremy Faden$^{1}$, Ivar Christopher$^{1}$, Chris Piker$^{1}$,
Sarah A. Henderson$^{1}$, Daniel Da Silva$^{3,5,6}$,
Scott Bounds$^{1}$, and Robert Strangeway$^{7,8}$
\par}

\vspace{0.3cm}

{\fontsize{10}{12}\selectfont
$^{1}$Department of Physics and Astronomy, University of Iowa, Iowa City, IA, USA\\
$^{2}$Southwest Research Institute, San Antonio, TX, USA\\
$^{3}$University of Texas, San Antonio, TX, USA\\
$^{4}$Space Sciences Laboratory, University of California, Berkeley, CA, USA\\
$^{5}$Solar Physics Laboratory, NASA Goddard Space Flight Center, Greenbelt, MD, USA\\
$^{6}$University of Maryland, Baltimore County, Baltimore, MD, USA\\
$^{7}$Laboratory for Atmospheric and Space Physics, University of Colorado, Boulder, CO, USA\\
$^{8}$University of California, Los Angeles, USA
\par}

\end{center}

\vspace{0.4cm}

\noindent\textbf{Corresponding author:} Shirsh Lata Soni, \href{mailto:shirshlata-soni@uiowa.edu}{shirshlata-soni@uiowa.edu}

\vspace{0.5cm}

\noindent\textbf{Key Points}
\begin{itemize}
    \item Cusp precipitation evolves across storm phases, with a poleward shift and transition from standard to reverse dispersion.
    \item Event-specific derived transit distances show a systematic increase, demonstrating evolving magnetic connectivity during storm evolution.
    \item Poleward cusp displacement is consistent with IMF-controlled reconnection geometry and shifts in the open-closed boundary.
\end{itemize}

\vspace{0.5cm}

\section{Abstract}
The dayside cusp provides a direct pathway for solar wind plasma entry into the magnetosphere--ionosphere system through magnetic reconnection. Using low-altitude ion and electron measurements from TRACERS, together with upstream solar wind and geomagnetic conditions, we investigate the evolution of the cusp during a geomagnetic storm on 30 September 2025, spanning its rising, main, and recovery phases, and compare these with a quiet-time reference. Storm-time observations show broader and more poleward precipitation regions and enhanced electron energy flux, indicating intensified dayside coupling. To interpret these variations, we combine solar wind and IMF measurements with the maximum magnetic shear reconnection model to determine X-line locations and use a Tsyganenko field model to compute event-specific field-line transit distances between the X-line and TRACERS. The results demonstrate that cusp morphology and latitude track IMF-driven reconnection geometry, and that realistic path lengths are essential for quantitative reconnection-rate estimates, highlighting the capability of TRACERS to resolve storm-time cusp evolution. Enhanced cusp precipitation during the recovery phase is consistent with IMF conditions, indicating sustained solar wind driving rather than intrinsic storm-phase effects.

\section{Plain Language Summary}
Geomagnetic storms reshape Earth’s magnetic shield and funnel solar‑wind particles into narrow regions near the poles called cusps. These cusps provide a direct window into how, where, and how strongly the solar wind reconnects with Earth’s magnetic field. In this study, we analyze three cusp crossings by NASA’s TRACERS mission during a single storm. We show that the storm‑time cusps become wider and move poleward, and that the incoming electron energy increases strongly, indicating more intense energy transfer from the solar wind to Earth’s upper atmosphere. To understand where this energy comes from, we use models to locate the magnetic reconnection line on the dayside boundary of Earth’s magnetic field and to trace magnetic field lines from this boundary down to the TRACERS spacecraft. These field‑line paths let us convert the particle measurements into estimates of the reconnection strength. Our results link the changing shape and position of the cusp directly to changes in the reconnection location and efficiency during the storm. This work demonstrates how TRACERS can track the evolving “geo-space weather” engine and sets up future studies using many more storms.

\section{Introduction}
Magnetic reconnection at Earth’s dayside magnetopause controls the entry of solar‑wind plasma and magnetic flux into the magnetosphere and is a primary driver of geomagnetic storms \cite{Dungey1961, Gonzalez1994}. During storm main phases, enhanced dayside reconnection and strong solar‑wind forcing substantially restructure the high‑latitude cusp, altering its location, width, and precipitation characteristics \cite{NewellMeng1987, Yamauchi1996}. However, most in situ cusp studies \cite{LockwoodSmith1994, Pitout2009} and storm time cusp studies \cite{Meng1982LatitudinalCuspStorm, Meng1983StormTimeCusp} have focused on isolated or moderately disturbed intervals, leaving the detailed evolution of cusp topology and precipitation across different storm phases, (rising, main, and recovery) poorly quantified. In particular, the response of cusp precipitation to evolving IMF conditions during geomagnetic storms remains incompletely quantified, particularly at low altitudes, limited observations, and what uncertainties arise when inferring reconnection properties from single‑spacecraft cusp crossings.

Several missions have provided important constraints on cusp morphology, including its typical latitudinal extent, energy‑latitude dispersion signatures, and relationships to interplanetary magnetic field (IMF) orientation and solar‑wind dynamic pressure \cite{NewellMeng1988, Newell2006, Newell2007}. Statistical models derived from these data capture average cusp locations, but they do not resolve rapid, storm‑time evolution or distinguish temporal changes from spatial structure along the cusp boundary \cite{CarbaryMeng1986, Zhou2000}. Multiple cusp crossing observations, especially in coordinated low‑altitude orbits, offer a powerful means to separate spatial and temporal variability by sampling closely spaced field lines under nearly identical upstream conditions \cite{Escoubet2001, Dunlop2002}. Such measurements are crucial for assessing how well cusp‑based reconnection diagnostics represent the global dayside reconnection geometry during strong geomagnetic activity \cite{Milan2007, Trenchi2008}.

In this study, we use observations from one of NASA’s Tandem Reconnection and Cusp Electrodynamics Reconnaissance Satellites (TRACERS) \cite{Miles2025_TRACERS_Mission, Petrinec2025_TRACERS_Design, Christopher2025_TRACERS_SOC} during the September 30, 2025, geomagnetic storm to probe storm‑time cusp dynamics. The spacecraft is in high‑inclination low‑Earth orbit (altitude=590 km) and repeatedly traverses the northern cusp and adjacent boundary layers, measuring low‑energy ions and electrons together with fields. In this study, we use observations from a single TRACERS spacecraft because tandem measurements were unavailable due to an anomaly affecting the TRACERS-1 during the interval analyzed. We focus on three cusp crossings that occur during the storm’s rising, main, and recovery phases, respectively, and compare them with a non‑storm baseline cusp pass from the preceding day. For each crossing, we combine upstream solar‑wind and SYM‑H measurements with TRACERS particle observations and a statistical cusp model to characterize cusp location, width, and precipitation signatures.

Recent work by \citep{daSilva2025CuspReconnection} showed that cusp ion‑energy dispersions can be used to estimate the dayside reconnection rate and that the inferred reconnection electric field is highly sensitive to the assumed transit distance between the reconnection site and the observing spacecraft. Building on this framework, we adopt a geometry‑aware approach that uses modeled reconnection line locations and magnetic field tracing to examine how storm‑time cusp precipitation observed by TRACERS reflects both the strength of reconnection and the evolution of the surrounding magnetic topology.

Our objective is to trace how the cusp responds to changing solar-wind and geomagnetic conditions during a single geomagnetic storm, assess the role of reconnection geometry and field-line path length in shaping the results, and establish a template for future storm-time, multi-event studies with TRACERS. By contrasting storm‑time cusp passes with a quiet‑time baseline, we examine how reconnection‑driven changes in cusp morphology and precipitation unfold from storm onset through peak and early recovery, and how these changes relate to modeled dayside reconnection locations and rates. This single‑spacecraft analysis provides observational constraints on storm‑time reconnection dynamics at the dayside magnetopause and lays the groundwork for statistical reconnection studies using the broader TRACERS cusp database.

\section{Methodology}
We first identified a September 30, 2025, storm and three TRACERS cusp crossings sampling the rising, main, and recovery phases, plus a quiet‑time baseline pass, using upstream solar‑wind/IMF data and SYM‑H. For each interval, we used TRACERS Analyzer for Cusp Ions (ACI) \cite{Fuselier2025ACI} and Analyzer for Cusp Electrons (ACE) \cite{Halekas2025TRACERSACE} differential energy‑flux data, constructed energy-latitude spectrograms, and consistently defined poleward boundary layer, core cusp, and equatorward boundary layer from soft‑ion wedges, ion dispersions, and accompanying electron signatures. Precipitating electron energy flux was then computed by converting ACE spectra to energy flux and integrating over the full angular distribution, taking the core‑cusp maximum for baseline and each storm crossing.

To relate cusp observations to reconnection geometry, we applied the Maximum Magnetic Shear model \cite{Trattner2021,Fuselier2022}, driven by IMF and solar-wind parameters convected from L1, to obtain the dayside reconnection X-line location for each crossing and compare it with the TRACERS cusp footprints. We then merged this X-line with a Tsyganenko-type field model \cite{Tsyganenko1995}: for X-line points near the spacecraft local time, field lines were traced Earthward, adjusted slightly inward if they were outside the model magnetopause, and the path length from the X-line to the spacecraft was taken as the event-specific transit distance $d^{\prime}$. Using these $d^{\prime}$ values and the observed low-energy ion cutoffs in the cusp, we applied the \citep{Lockwood1992} method to derive along-track reconnection electric fields for each crossing. Finally, we quantified the latitudinal extent and centroid of ion and electron precipitation and compared these with modeled X-line motion and inferred reconnection-rate variations to assess storm-time control of cusp structure. This approach assumes that changes in ion-energy dispersion primarily reflect temporal variations in reconnection; however, such temporal evolution cannot be uniquely determined from single-spacecraft measurements and may also include spatial effects.

\section{Observation and Results}
\subsection{  Storm‐time context and scientific motivation}

 Storm‑time cusp intervals provide a particularly sensitive probe of how solar‑wind driving and magnetopause reconnection respond to rapidly changing upstream conditions. When the IMF turns southward and solar‑wind dynamic pressure increases, enhanced dayside reconnection is expected to both intensify cusp precipitation and modify the latitude and structure of the cusp footprint. Capturing multiple cusp traverses (every 90 minutes) by the same spacecraft during one storm, therefore, offers a rare opportunity to follow this evolution in a controlled way, using identical instrumentation and nearly identical orbital geometry. 

\begin{figure}
 \includegraphics[width=\textwidth]{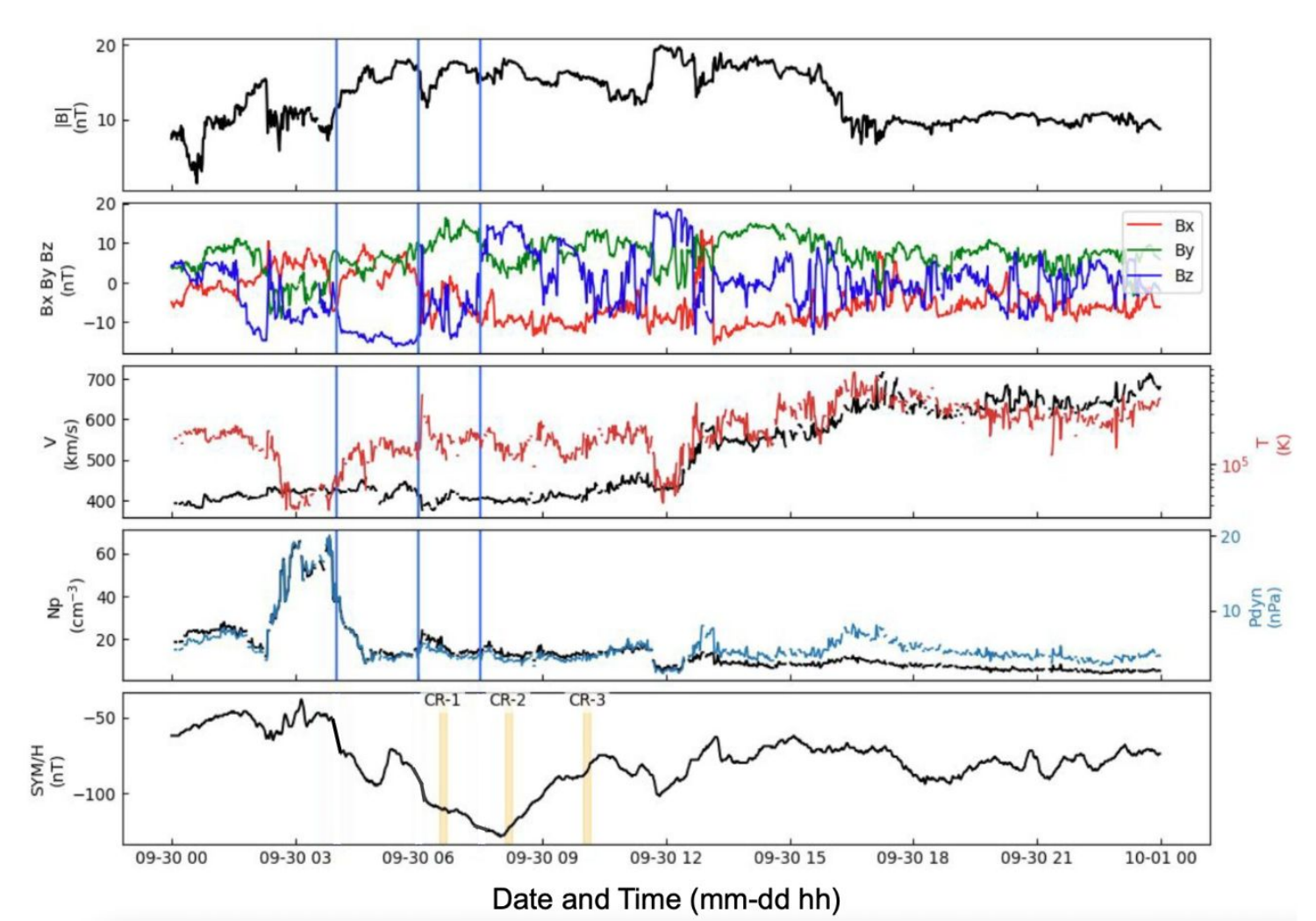}
 \caption{Upstream solar wind and geomagnetic conditions during the 30 September storm. From top to bottom, the panels show the interplanetary magnetic field magnitude, IMF components (Bx, By, Bz), solar wind proton density, temperature, bulk speed, dynamic pressure, and the SYM-H index. The vertical yellow lines mark the times of the three TRACERS cusp crossings (CR-1, CR-2, and CR-3), corresponding to the storm rising, main, and recovery phases, respectively. Vertical blue lines show the initial IMF parameters time-shifted from L1; vertical lines mark cusp crossing times, not direct correspondence. These parameters provide the large-scale solar wind and geomagnetic context for the storm-time cusp observations analyzed in this study.} 
 \label{Fig1}
 \end{figure}

Figure \ref{Fig1} summarizes the upstream solar‑wind conditions for the 30 September storm. The three TRACERS cusp crossings analyzed in this study occur successively during the storm’s rising, main, and early recovery phases and are marked by the vertical bands in the bottom panel along with geomagnetic conditions (Sym-H), providing a natural sequence with which to examine how cusp morphology and reconnection signatures evolve under systematically changing upstream driving. Data immediately prior to the storm-time cusp crossing and during 29 September were not available, and therefore those intervals are not included in the present analysis. The IMF magnitude remains elevated, and the GSM Bz component repeatedly turns southward to values near -10 nT, creating intervals of strong dayside coupling. A relatively high and steady solar‑wind speed of order ${500 kms^{-1}}$ implies large poynting flux into the magnetosphere, while superposed density enhancements drive dynamic‑pressure pulses that can compress the magnetopause and further intensify reconnection. The geomagnetic response, as indicated by SYM‑H, exhibits a well‑defined main‑phase decrease followed by gradual recovery. 

\subsection{ Storm‐time cusp crossings and morphology}
\begin{sidewaysfigure}
    \centering
    \includegraphics[width=1\linewidth]{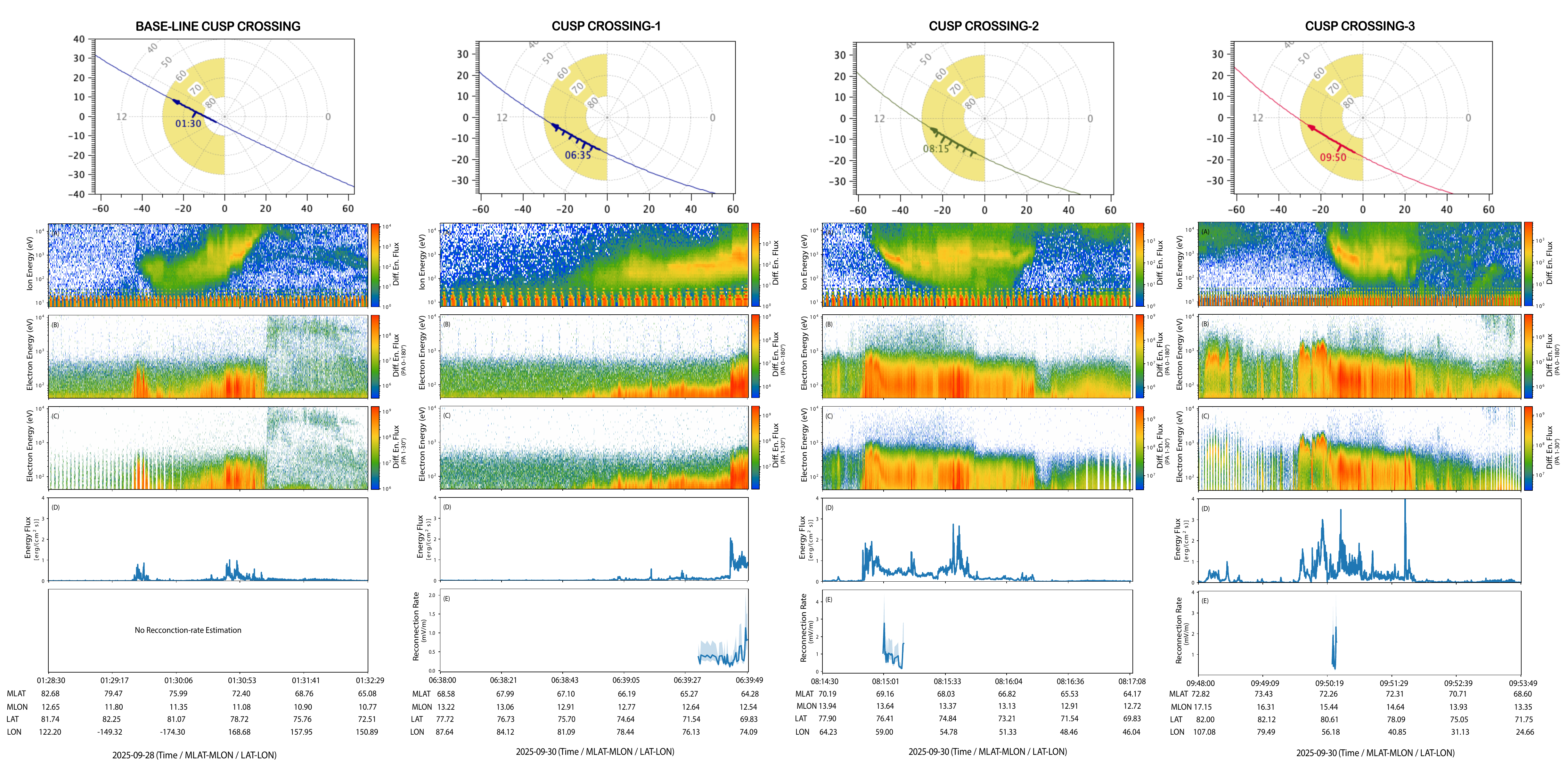}
    \caption{Spatial context and particle precipitation observed by TRACERS during baseline and storm-time cusp crossings. Columns show (from left to right) a non-storm baseline interval and three cusp crossings during the storm rising (CR-1), main (CR-2), and recovery (CR-3) phases. The top panels show the spacecraft trajectories in magnetic coordinates. The dashed circles indicate magnetic latitude, and the shaded regions highlight the cusp location. (colored lines). The middle panels show ion (a) and electron (b) energy spectrograms, illustrating enhanced and broadband particle precipitation during storm-time cusp crossings compared to baseline conditions, and (c) shows the precipitating electron energy fluxes integrated over pitch angles 0-30°. Bottom panels present the corresponding particle energy flux, highlighting storm-phase-dependent intensification of cusp precipitation (d). An additional panel showing reconnection rate estimates will be included (e).}
    \label{Fig2}
\end{sidewaysfigure}

Figure \ref{Fig2} summarizes the morphology of the three storm‑time cusp crossings; each panel combines the spacecraft ground track over a northern pole with ACI ion and ACE electron spectrograms, allowing us to define poleward and equatorward boundary layers (BLs) and the cusp in a consistent way. 

Crossing 1 (CR-1) (06:38:30–06:40:30 UT) occurs during the rising phase of the storm and is characterized by a being within a limited latitudinal range ($\sim2–3^\circ$ MLAT), with weak ion dispersion and modest electron flux. In the polar cap, ACI detects weak, patchy low-energy precipitating ions, while core‑cusp signatures are confined to a narrow interval near 06:39:00–06:39:50 UT with the maximum soft‑ion flux of the pass but only limited ion dispersion; the equatorward BL is not observed in this interval. 

Crossing 2 (CR-2) (08:13:30–08:17:10 UT), near the storm's main phase, shows a much more developed cusp.  The poleward BL contains the first appearance of low-energy ions with emerging reverse dispersion, the core cusp exhibits strong magneto-sheath‑like ions in a smooth broadband wedge together with a clear step-like cusp signature, and the equatorward BL shows structured high‑energy ions and decaying soft‑electron flux. This interval occurs under predominantly northward IMF conditions, consistent with a transition to high-latitude reconnection and the observed reverse ion dispersion.

Crossing 3 (CR-3) (09:48:00–09:51:45 UT), in the early recovery phase, samples a broad cusp embedded between auroral‑zone and polar‑cap precipitation. The poleward BL shows auroral signatures and multiple inverted‑V electrons, the core cusp contains a very strong, broad soft‑ion wedge with pronounced reverse dispersion and enhanced, somewhat softer electrons, and the equatorward BL is marked by patchy spectra and sharp decreases in ion and electron flux. These three passes thus provide a sequence of storm‑time cusp states, from weakly dispersed to broad, intense, and structured cusp, that we relate in later sections to changes in reconnection geometry and rate. The persistence of broad and intense precipitation during this interval is consistent with continued favorable IMF (southward Bz for CR-1: day side reconnection and northward Bz for CR-2, CR-3:  high latitude (lobe) reconnection) conditions rather than storm-phase effects alone.

In addition to the three storm-time passes, we analyze a quiet-time “baseline” cusp crossing from the preceding day, selected under relatively steady solar-wind conditions and near-zero SYM-H. During this interval, the spacecraft traverses a well-defined cusp confined to a limited latitudinal extent, characterized by modest soft-ion fluxes and clear ion energy–latitude dispersion, but lacking strong reverse dispersion signatures. The electron spectra vary gradually with latitude and exhibit minimal temporal or spatial variability. The poleward and equatorward boundary layers are both present but occupy a smaller latitudinal range than during the storm, and the overall morphology is more spatially uniform, with reduced variability and fewer distinct spectral features. This baseline pass provides a reference for typical non-storm cusp location, width, and precipitation characteristics, against which the storm-time changes in intensity and geometry in Crossings 1–3 can be quantitatively assessed.

\subsection{ Electron energy flux in storm‑time cusps}
To quantify how storm‑time conditions modify cusp precipitation, we derived precipitating electron energy fluxes integrated over all anergy and angles. For each interval (baseline and CR‑1/2/3), differential energy flux spectra from TRACERS were integrated over energy, yielding a single representative peak value for the core‑cusp segment of each pass. This procedure ensures a consistent comparison between the non-storm‑time reference cusp and the three storm‑time crossings.

The resulting maximum electron energy fluxes (in ergs cm$^{-2}$ s$^{-1}$) show a clear enhancement during the geomagnetic storm. The quiet‑time baseline cusp exhibits a peak value of approximately 1, while the storm‑time crossings reach  $\sim$2 for CR‑1,  $\sim$3 for CR‑2, and  $\sim$3.5 for CR‑3. Thus, even during the storm's rising phase (CR‑1), the field‑aligned electron energy flux is roughly doubled relative to the non‑storm cusp (based on the given limited observation). The main‑phase crossing (CR‑2) shows a threefold increase, and the recovery‑phase crossing (CR‑3) retains an elevated level, exceeding the baseline by a factor of  $\sim$3.5 despite partial relaxation of the global geomagnetic disturbance.

\subsection{ Reconnection location from the Maximum Magnetic Shear model}
\begin{figure}

 \includegraphics[width=\textwidth]{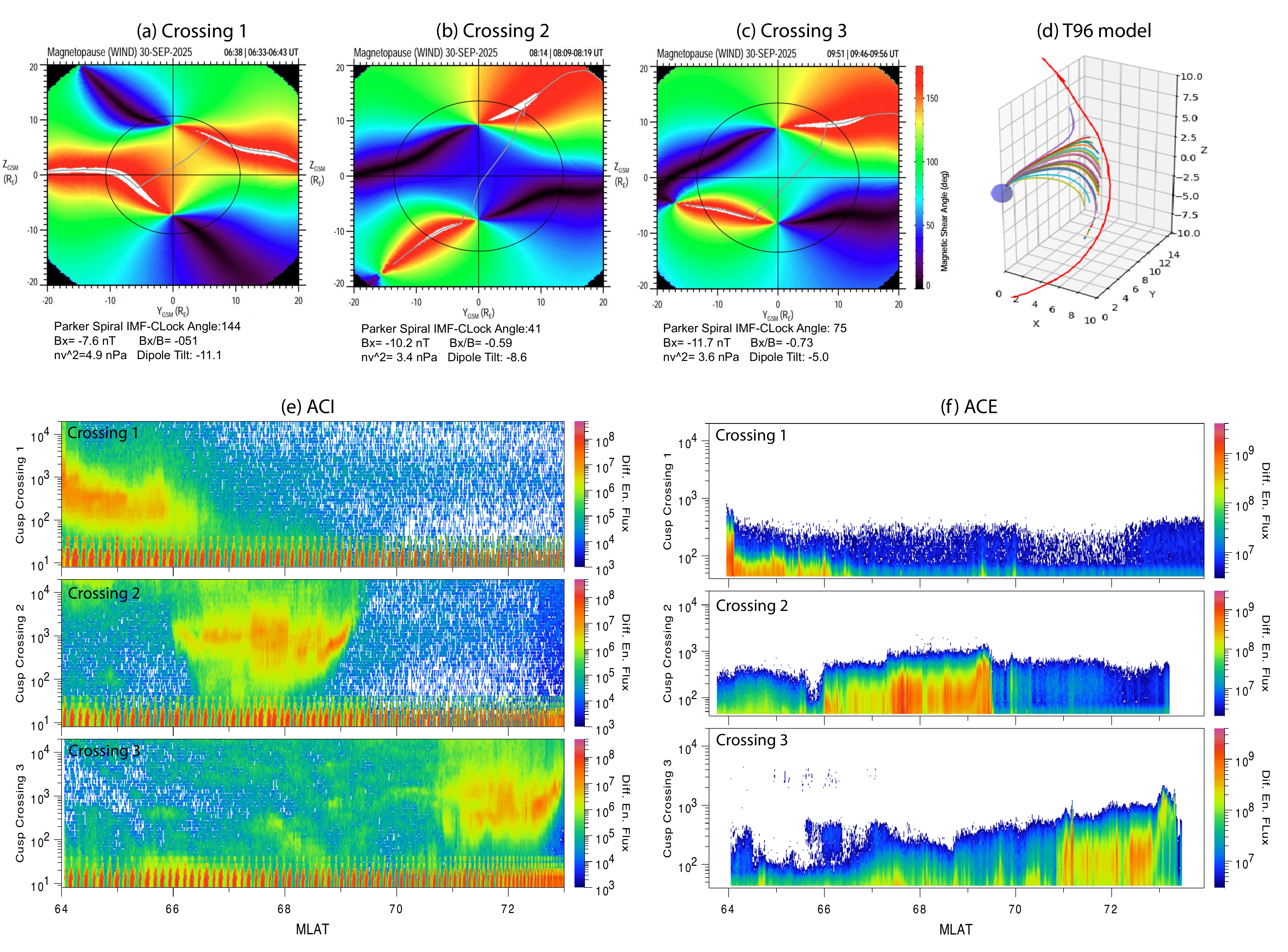}
 \caption{Storm‑time cusp evolution during the September 30, 2025, geomagnetic storm. Top row: Maximum magnetic shear maps at the dayside magnetopause for the three TRACERS cusp crossings (a-c: Crossings 1–3), showing the modeled reconnection line (white), IMF clock angle, solar‑wind dynamic pressure, and dipole tilt at the time convected to the magnetopause. The 3‑D panel (d) illustrates representative T96 field‑line traces from the maximum‑shear X-line to the TRACERS spacecraft, used to derive event‑specific transit distances. (e): ACI ion differential energy‑flux spectrograms versus magnetic latitude (MLAT) for the three crossings, highlighting the evolution from a thinner, weakly dispersed cusp in Crossing 1 to broader, more structured soft‑ion wedges in Crossings 2 and 3. (f)): Corresponding ACE electron spectrograms versus MLAT, showing the concurrent changes in electron precipitation, including enhanced and poleward‑shifted fluxes during the main and recovery phases.}
 \label{Fig3}
 \end{figure}

To examine where dayside reconnection occurred for each cusp interval, we applied the empirical Maximum Magnetic Shear (MMShear) model of Trattner et al. (2007, 2021). For each crossing, upstream IMF and solar‑wind conditions were time‑shifted from L1 to the cusp using the observed solar‑wind speed, giving cusp arrival delays of 1 h 01 min (CR‑1), 1 h 39 min (CR‑2), and 1 h 41 min (CR‑3) (Figure \ref{Fig3}). These convected inputs were then used, together with the T96 internal field and a draped magnetosheath field model \cite{Tsyganenko1995} to compute the magnetic shear angle across the dayside magnetopause and to trace the predicted reconnection line on the magnetopause surface.

For the rising‑phase cusp crossing CR‑1, the IMF clock angle was about 144°, defined as $\theta_c = \tan^{-1}(B_y/B_z)$ in GSM coordinates, was $\sim 144^\circ$, corresponding to strongly southward IMF with a modest By component. Under these conditions, the MMShear model places the reconnection X-line along an extended anti‑parallel branch near the subsolar region, consistent with classical southward‑IMF reconnection and limited ion dispersion observed by TRACERS. In contrast, during the main‑phase crossing CR‑2, the clock angle decreased to $\sim$41°, indicating dominantly northward IMF with significant By. The MMShear solution then shifts the reconnection line poleward into the high‑latitude dayside magnetopause, where large shear persists along a tilted, quasi‑continuous X-line poleward of the northern cusp, in agreement with the strong, smooth reverse ion dispersion and step‑cusp signatures in the TRACERS data. This interval corresponds to northward IMF conditions, consistent with high-latitude reconnection and the observed reverse ion dispersion.

By the recovery‑phase crossing CR‑3, the IMF clock angle had rotated further to $\sim$75°, again favoring high‑latitude, predominantly northward reconnection. The predicted X-line lies even closer to the cusp throat, along a high‑shear band that maps directly to the latitudes where TRACERS observes broad soft‑ion wedges and well‑developed reverse dispersions. The persistence of broad precipitation during this interval is consistent with continued favorable IMF conditions rather than storm-phase effects alone. In all three cases, the separation between the spacecraft ground track and the modeled X-line is within a few Earth radii, comparable to the stated $\sim$2 $R_E$ uncertainty of the MMShear model. This consistency supports the interpretation that the observed storm‑time cusp signatures arise from reconnection occurring along the maximum‑shear locus whose latitude and local‑time position evolve with IMF clock angle and storm phase, rather than from a fixed subsolar X-line. The persistence of broad and intense precipitation during this interval is consistent with continued favorable IMF conditions rather than storm-phase effects alone.

\subsection{ Reconnection rate from cusp ion cutoffs with updated transit distances}
We next quantified the dayside reconnection rate for each crossing using the Lockwood (1992) method, which relates the low‑energy cutoff in cusp ion dispersions to the reconnection electric field at the magnetopause. In this framework, the reconnection rate depends linearly on the distance $d'$ that precipitating ions travel along newly opened field lines between the reconnection site and the cusp observer; accurate estimation of $d'$ is therefore essential. To obtain storm‑time appropriate transit distances, we combined the T96 geomagnetic field model with the Maximum Magnetic Shear X‑line locations of \cite{Trattner2007, Trattner2021}, using upstream IMF and solar‑wind conditions convected from L1 to the magnetopause as in Section 3.4. Magnetic shear is defined as the angle between the magnetosheath and magnetospheric magnetic fields across the magnetopause. The maximum magnetic shear model identifies the reconnection X-line as the locus where this angle is maximized. We note that the derived values represent local reconnection electric fields at specific locations along the X-line. A global reconnection rate would require integration along the entire X-line and cannot be determined from single-spacecraft observations.

For each event, we first selected the segment of the Trattner X-line within a narrow local‑time window around the TRACERS footprint. From each candidate X‑line point, we traced field lines inward using T96, checking whether the starting position lay inside the model magnetopause; if not, the starting point was moved radially inward by 2.5\% and the trace repeated until it did. Among all such traces, we then identified the field line that passed closest to the spacecraft position in GSM coordinates and defined its length between the X‑line and the spacecraft as the effective Lockwood transit distance $d'$. This procedure yields transit distances of 7.95 $R_E$ for CR-1, 12.25 $R_E$ for CR-2, and 15.49 $ R_E$ for CR-3, reflecting the increasing separation between the reconnection line and the cusp footprint as the storm progresses into main and recovery phases.

Using these event‑specific $d'$ values together with the observed low‑energy cutoffs in the TRACERS ACI ion spectra, we recomputed the Lockwood reconnection rate profiles along each cusp pass. The resulting rates reach values of order 1-2 mV/m in localized peaks, with higher and more sustained reconnection rates during the main‑phase crossing (CR-2) than during the rising‑phase crossing (CR-1), and narrow, intense bursts during the recovery‑phase crossing (CR-3). Compared to estimates that assume a fixed or purely model‑based transit distance, the new rates are systematically reduced for the longer‑path events yet remain consistent with strong storm‑time dayside reconnection, indicating that part of the apparent event‑to‑event variability in previous cusp‑based estimates may arise from oversimplified assumptions about $d'$. This merged T96-maximum‑shear approach thus provides a self‑consistent way to couple TRACERS cusp observations to realistic reconnection geometries and yields more robust constraints on the magnitude and temporal structure of the dayside reconnection electric field during geomagnetic storms.

\subsection{ Latitudinal evolution of ion and electron precipitation}
The TRACERS ACI (ions) and ACE (electrons) spectrograms for the three storm‑time crossings reveal a systematic poleward shift and broadening of the cusp precipitation region as the storm evolves (Figure \ref{Fig3}). In CR-1 (early phase, top panels), enhanced soft‑ion and electron fluxes are confined within approximately 65-68° magnetic latitude, with weak or patchy flux poleward of the core precipitation. In CR-2 (main phase, middle panels), the region of intense low‑energy ions and electrons expands both equatorward and poleward, filling roughly 66-71° and exhibiting strong energy‑latitude dispersion in the ions and a pronounced softening of the electron spectrum toward higher latitudes. By CR-3 (recovery phase, bottom panels), the brightest precipitation shifts further poleward, with structured, intermittent flux extending to $\sim$72-73°, while equatorward flux is reduced compared to the main phase.

This latitudinal evolution is consistent with the maximum‑shear reconnection geometry and the Lockwood model‑based reconnection rates derived earlier. During the rising phase (CR-1), the shorter modeled transit distance (7.95 $R_E$) and thin cusp (Sections 3.4-3.5) correspond to a relatively low, narrowly localized reconnection rate and a cusp footprint confined to lower latitudes. At the main phase (CR-2), the transit distance increases to 12.25 $R_E$, and the Lockwood rates reach 1-2 mV/m, in tandem with a substantial widening and poleward shift of both ion and electron precipitation, indicating that strong dayside reconnection maps to a broader cusp throat. In the recovery phase (CR-3), the longest transit distance (15.49 $R_E$) and bursty reconnection signatures coincide with a cusp footprint that has migrated to still higher latitudes and become more structured, while the electron energy flux remains elevated relative to the non‑storm baseline (Section 3.3).

These results are consistent with storm-time conditions enhancing cusp energy deposition relative to baseline conditions. The elevated energy flux during CR-2 and CR-3 reflects increased particle fluxes in the magnetosheath, driven by solar wind conditions, which enhance the supply of particles available for precipitation into the cusp. The cusp precipitation responds directly to upstream conditions and may not track global geomagnetic indices such as SYM-H on short timescales.

\section{Discussion and Conclusions}
\label{sec:discussion}

The September 30, 2025, storm provides a rare sequence of three cusp crossings by the same TRACERS spacecraft (facilitated by its ~90-minute orbit) during the rising (CR-1), main (CR-2), and early recovery (CR-3) phases, together with a nearby quiet-time baseline pass. This configuration allows a systematic comparison of cusp morphology and precipitation under different levels of solar wind driving, using identical instrumentation and nearly identical orbital geometry. By examining low-energy ions (ACI) and electrons (ACE) together with upstream solar wind and IMF data, we follow the cusp response from a thin, weakly dispersed structure early in the storm to a broad, intense cusp at storm peak, and then to a structured, poleward-shifted
cusp during recovery.

To interpret the storm‑time cusp changes in a physically consistent way, we also relate the low‑altitude observations to the expected reconnection geometry at the dayside magnetopause. For each crossing, we use the maximum‑magnetic‑shear model to estimate the location of the reconnection line and combine this with a global magnetic‑field model to obtain an event‑specific field‑line transit distance between the X-line and the spacecraft. This distance enters the Lockwood reconnection‑rate calculation based on cusp ion low‑energy cutoffs, allowing the inferred rates to reflect the actual field‑line paths for each storm phase. In this framework, the observed variations in cusp latitude, width, and spectra can be understood in terms of changes in both local precipitation and the larger‑scale reconnection geometry.
On this basis, our main quantitative conclusions are:

\begin{itemize}
  \item[1] Storm-time reconnection substantially enhanced cusp energy input relative to quiet conditions.
    \begin{itemize}
      \item Peak electron energy flux in the core cusp increases
      from about 1 erg~cm$^{-2}$~s$^{-1}$ in the baseline to roughly 2, 3, and
      3.5~erg~cm$^{-2}$~s$^{-1}$ in CR-1, CR-2, and CR-3, respectively, implying a factor
      of $\sim 2$-3.5 enhancement in electron energy input along cusp field lines.
      \item The enhancement during CR-1 occurs under southward IMF $(B_z < 0)$, consistent with strong dayside reconnection, whereas the elevated fluxes during CR-2 and CR-3 occur under northward IMF conditions and are associated with high-latitude (lobe) reconnection and continued magnetosheath particle access to the cusp.
    \end{itemize}

  \item[2] Cusp morphology and latitude tracked the evolving reconnection geometry at the magnetopause.
    \begin{itemize}
      \item The magnetic latitudinal width of intense ion/electron precipitation grows from a narrow
      ($\sim 2^\circ$-$3^\circ$) band in CR-1 to a broader ($\sim 4^\circ$-$5^\circ$) cusp
      in CR-2, and remains wide but more structured in CR-3, while the centroid latitude
      shifts poleward by several degrees between CR-1 and CR-3.
      \item Maximum-shear modeling shows the reconnection X-line moving from lower-latitude,
      southward-IMF geometry (clock angle $\sim 144^\circ$) during CR-1 to higher-latitude,
      more northward geometry (clock angles $\sim 41^\circ$ and $\sim 75^\circ$) during
      CR-2 and CR-3, matching the observed poleward motion and broadening of cusp precipitation.
    \end{itemize}

  \item[3] Realistic field-line path lengths were essential for quantitative cusp-based reconnection rates.
    \begin{itemize}
      \item Event-specific transit distances $d'$ between the X-line and \textit{TRACERS},
      obtained by tracing field lines from the maximum-shear line, increases from
      7.95~$R_E$ (CR-1) to 12.25~$R_E$ (CR-2) and 15.49~$R_E$ (CR-3), reflecting
      progressively more distant reconnection geometries.
      \item When these $d'$ values are used in the Lockwood (1992) formulation, inferred
      reconnection electric fields along the cusp reach localized peaks of order
      1-2~mV/m during CR-2 and CR-3, remaining strong but avoiding unrealistically
      large values that would result from assuming a single, fixed path length.
      \item Theoretically, because $E_R$ in cusp cutoff methods scales inversely with the assumed path length, using too short a distance can artificially inflate reconnection rates; the event‑specific $d'$ used here mitigates this bias and produces values consistent with MHD and kinetic estimates for dayside reconnection under similar solar‑wind conditions.
    \end{itemize}

  \item[4.] Cusp precipitation during the recovery phase reflected continued IMF-controlled reconnection.
    \begin{itemize}
      \item During CR-3, cusp electron energy flux remained elevated relative to quiet-time conditions, and ion spectra continue to show clear reverse dispersions despite partial recovery in SYM-H.
      \item This behavior is consistent with continued IMF variability, including transitions from southward to northward $B_z$, rather than a delayed magnetospheric response.
      \item These observations are consistent with cusp energetics being primarily controlled by solar wind driving rather than global geomagnetic indices on short timescales.
    \end{itemize}

\end{itemize}

Together, these observations demonstrate how cusp morphology and reconnection efficiency evolve throughout a storm, establishing TRACERS as a powerful platform for storm‑time cusp and reconnection studies. By analyzing closely spaced, near‑simultaneous traversals of the two spacecraft, it will be possible to distinguish features that evolve in time along a single flux tube (such as pulsed reconnection or transient flux ropes) from persistent gradients across neighboring flux tubes (such as localized X‑line segments or large‑scale shear in the boundary). Combining this separation with along‑track reconnection‑rate estimates will constrain the characteristic spatial scales and intermittency of storm‑time reconnection at the magnetopause. In parallel, coupling TRACERS cusp diagnostics with conjugate ground‑based observations and with global magnetohydrodynamic or kinetic simulations will embed the storm‑time evolution of cusp topology within a full Sun–magnetosphere–ionosphere framework, enabling tests of how changes in reconnection geometry map into ionospheric electrodynamics and thermospheric energy deposition.

This study provides a phase-resolved, low-altitude observational constraint on IMF-controlled cusp dynamics during geomagnetic storms, highlighting the capability of TRACERS to directly link magnetopause reconnection geometry with ionospheric energy deposition.

\section{Data and Software Availability}
TRACERS ion (ACI) and electron (ACE) data used in this study are available from the mission science data archive (https://tracers-portal.physics.uiowa.edu/). Upstream solar wind and IMF data were obtained from the OMNI database provided by NASA’s Space Physics Data Facility (https://omniweb.gsfc.nasa.gov). Geomagnetic indices (e.g., SYM-H/Dst) were also obtained from OMNI.

\section{Acknowledgments}
We thank the TRACERS instrument and operations teams for providing the high‑quality data products that made this study possible. We acknowledge the use of solar wind and geomagnetic indices from space‑weather data centers and thank the developers and maintainers of the magnetic field and cusp models employed in our analysis. The TRACERS mission is supported by NASA through contract 80GSFC18C0008.

\bibliographystyle{plain}
\bibliography{references} 
%

\end{document}